\documentclass[twocolumn,aps,showpacs,prb,tightenlines,amsmath,amssymb,superscriptaddress]{revtex4}
\usepackage{graphicx}

\usepackage{dcolumn}

\begin{document}

\title{Scanning Raman spectroscopy of graphene antidot lattices: Evidence for systematic p-type doping}
\author{S.\ Heydrich}

\affiliation{Institut
f\"ur Experimentelle und Angewandte Physik, Universit\"at
Regensburg, D-93040 Regensburg, Germany}
\author{M.\ Hirmer}
\affiliation{Institut f\"ur Experimentelle und Angewandte Physik,
Universit\"at Regensburg, D-93040 Regensburg, Germany}
\author{C.\ Preis}
\affiliation{Institut f\"ur Experimentelle und Angewandte Physik,
Universit\"at Regensburg, D-93040 Regensburg, Germany}
\author{T.\ Korn}
\affiliation{Institut f\"ur Experimentelle und Angewandte Physik,
Universit\"at Regensburg, D-93040 Regensburg, Germany}
\author{J.\ Eroms}
\affiliation{Institut f\"ur Experimentelle und Angewandte Physik,
Universit\"at Regensburg, D-93040 Regensburg, Germany}
\author{D.\ Weiss}
\affiliation{Institut f\"ur Experimentelle und Angewandte Physik,
Universit\"at Regensburg, D-93040 Regensburg, Germany}
\author{C.\ Sch\"uller}
\email{christian.schueller@physik.uni-regensburg.de}
\affiliation{Institut f\"ur Experimentelle und Angewandte Physik,
Universit\"at Regensburg, D-93040 Regensburg, Germany}

\date{\today}

\begin{abstract}
We have investigated antidot lattices, which were prepared on
exfoliated graphene single layers via electron-beam lithography
and ion etching, by means of scanning Raman spectroscopy. The peak positions, peak widths and intensities of the characteristic phonon modes of the carbon lattice have been studied systematically in a series of samples. In the patterned samples, we found a systematic stiffening of the G band mode, accompanied by a line narrowing, while the 2D mode energies are found to be linearly correlated with the G mode energies. We interpret this as evidence for p-type doping of the nanostructured graphene.
\end{abstract}
 \maketitle
\newpage

Since the first report about the preparation of graphene single layers via mechanical exfoliation technique \cite{Geim2007}, the interest in this seemingly ideal two-dimensional system has grown enormously. For the most part, this is motivated by the vision that the two-dimensional carbon system is potentially a promising candidate for future electronic devices \cite{Han2007}. Scanning Raman spectroscopy has proven to be a powerful technique for the characterization and investigation of graphene samples. It enables a fast and nondestructive investigation of the electronic and structural properties of layered and laterally-structured samples with  sub-micron spatial resolution. Via Raman spectroscopy, e.g., one is able to identify the number of layers in a sample \cite{Ferrari2006,Malard2007}, to characterize the edge chirality \cite{You2008} or to probe strain \cite{Metzger2010}. Most importantly for this work, it was shown that the doping in graphene can be monitored by Raman spectroscopy \cite{Pisana2007,Yan2007}, even in the case of unintentional doping due to charged impurities \cite{Casiraghi2007,Stampfer2007,Casiraghi2009}.
The most prominent phonon modes in two-dimensional carbon lattices are: (i) The G band, which stems from inplane LO phonons with $E_{2g}$ symmetry at the center of the Brillouin zone, and which is centered around 1580 $\rm cm^{-1}$ in intrinsic graphene. (ii) The D mode (around 1340 $\rm cm^{-1}$), which is attributed to inplane TO phonons from around the K or K' points of the Brillouin zone, and which is forbidden due to wave-vector conservation. It requires a defect-induced scattering process to be observable, and appears in  high-quality graphene layers at the edges, only \cite{Casiraghi2009a}. (iii) The 2D mode, which is an overtone of the D mode. Due to the two-phonon nature of this mode, it is wave-vector conserving.
Finally, (iv), the D' band, which is again thought to be a defect-induced mode from the maximum of the LO phonon dispersion between $\Gamma$ and K. In this work, we have investigated systematically the Raman modes in a series of graphene samples, on which periodic hole lattices, i.e., so called antidot lattices, were prepared by electron-beam lithography and subsequent ion etching. Recently, in transport experiments on similar structures two of the authors observed pronounced weak localization due to strong intervalley scattering at the antidot edges and found a transport gap by analyzing thermally activated transport around the Dirac point.\cite{Eroms2009} Here, we find in our Raman experiments a strong correlation of the energetic positions of the G modes and 2D modes in all antidot samples, from which we conclude that the nanoscale patterning produces a p-type doping. We have found doping levels in the range of $3-7\times 10^{12}$ cm$^{-2}$.

\begin{figure}
  \includegraphics[width= 0.35\textwidth]{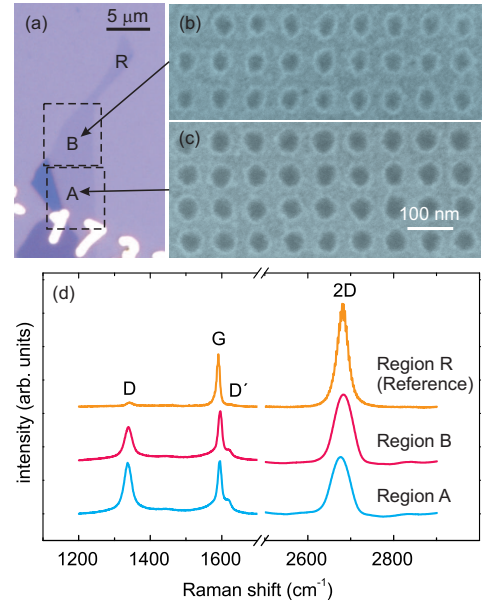}
   \caption{(a) Microscope image of a graphene flake. The dashed rectangles mark the regions, where antidot lattices were prepared.
   (b) Scanning electron micrograph of region B, and, (c) of region A. (c) Representative Raman spectra of regions A, B, and R on the sample.}
   \label{Fig1}
\end{figure}

\begin{figure}
  \includegraphics[width=0.4\textwidth]{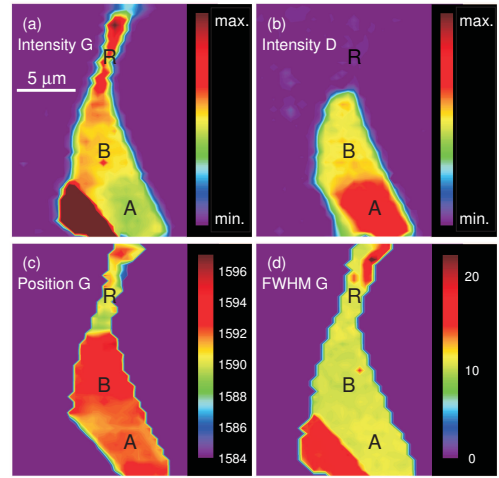}
   \caption{False color plots of scanning Raman experiments on the same sample as in Fig.\ 1. Displayed are measured images of (a) the intensity of the G mode, (b) the intensity of the D mode, (c) the energetic position of the G mode, and, (d) the linewidth of the G mode. The numbers at the scale bars in (c) and (d) are given in cm$^{-1}$.}
   \label{Fig2}
\end{figure}

\begin{figure}
  \includegraphics[width= 0.4\textwidth]{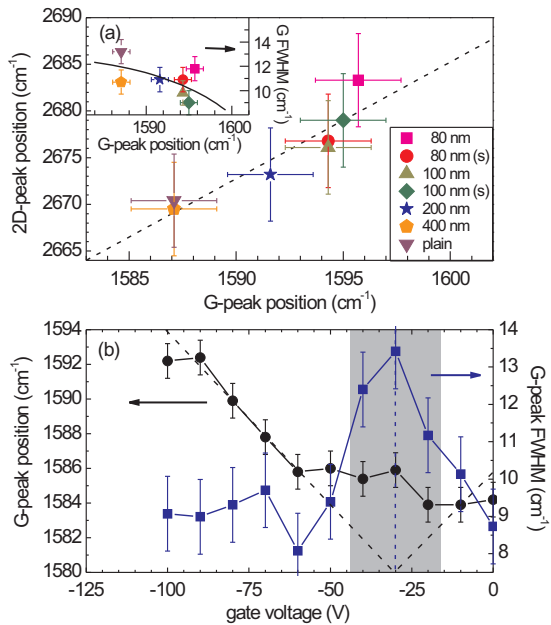}
   \caption{(a) Plot of the energetic positions of the observed 2D modes versus the positions of the G modes of all investigated antidot samples. The legend indicates the periods of the investigated antidot lattices. The diameter of the holes is approximately 60 nm. The label (s) means slightly smaller holes ($\sim $ 50 nm). The inset shows the linewidths of the G modes versus energetic position. (b) Plot of the energetic positions of the G mode and its linewidth versus gate voltage, as measured in a plain, gated graphene sample.}
   \label{Fig3}
\end{figure}

The graphene samples were prepared on Si wafers with 300 nm SiO$_{\rm 2}$ cap layers by the well-known mechanical exfoliation technique \cite{Geim2007}. On selected single layer flakes, antidot lattices with periods of 80 nm, 100 nm, 200 nm, and 400 nm were prepared using electron beam lithography and oxygen reactive ion etching. The hole diameters are between about 50 nm and 60 nm. In this series of samples, we have intentionally deposited no metal contacts in order to avoid unwanted additional doping due to the presence of metals on the samples \cite{Casiraghi2007}. We use below, however, measurements on a gated graphene reference sample for estimates of the hole densities. The Raman experiments were performed at room temperature with a 532 nm diode-pumped solid state laser and a microscope setup. The samples were mounted on a motorized x-y translation stage with minimal step sizes of 100 nm, the laser-spot diameter was 800 nm. Raman spectra were recorded with a TriVista triple Raman system, equipped with a nitrogen-cooled CCD detector. Figure \ref{Fig1}a shows an optical microscope image of one of the investigated samples. In the regions A and B, indicated by dashed rectangles in Fig.\ \ref{Fig1}a, antidot lattices with a period of 80 nm were prepared. The holes in region A have a slightly larger diameter ($\sim 60$ nm, see Fig.\ \ref{Fig1}c) than the holes in region B ($\sim 50$ nm, see Fig.\ \ref{Fig1}b). Region R is an un-patterned part of the sample. Note that on the left edge of region A and on the upper right edge of the reference region R, there are multi-layer areas, visible by the much darker contrast in the microscope image (Fig.\ \ref{Fig1}a). Typical Raman spectra of the three regions, A, B and R, are displayed in Fig.\ \ref{Fig1}d. In the antidot regions, A and B, pronounced D peaks are observable, along with D' peaks. We interpret this to be dominantly caused by the introduction of additional edges into the sample due to the creation of the holes, and not by additional defects or disorder, which might be caused by the preparation process. The reason for this interpretation is evident from the scanning Raman images, shown in Fig.\ \ref{Fig2}. Figure 2 shows false color plots of data, obtained from scanning Raman experiments on the
same sample, as shown in Fig.\ \ref{Fig1}. In these measurements, individual Raman spectra were taken with lateral spatial resolution of about 800 nm, in steps of 500 nm in x and y directions. From the raw spectra, peak positions, intensities and linewidths of the phonon modes were extracted. The image of the G mode intensity (Fig.\ \ref{Fig2}a) essentially reproduces the topology of the whole flake. The intensity of the G mode increases from region A, which contains the larger holes, to region B, with smaller holes, to the un-patterned reference region R. A maximum intensity is visible in the multi-layer part, left of region A. On the other hand, a pronounced intensity of the D mode is observed in the antidot areas A and B, only, with higher intensities in A, which contains the larger holes, i.e., with longer edges (Fig.\ \ref{Fig2}b). Interestingly, the energetic position of the G mode is systematically higher in the antidot lattices than in the reference part (see Fig.\ \ref{Fig2}c). While the G mode energy in the reference part is between about 1589 and 1590 cm$^{-1}$, it is in regions A and B upshifted to about 1592 to 1594 cm$^{-1}$. Evidence for the high crystalline quality of the prepared samples can be deduced from the image of the linewidths of the G mode, displayed in Fig.\ \ref{Fig2}d. The linewidths are over the whole single-layer region between about 9 and 11 cm$^{-1}$, comparable to pristine graphene samples \cite{Casiraghi2007, Casiraghi2009}, and to plain gated graphene outside the Landau-damping region (see below Fig.\ \ref{Fig3}b). For significantly disordered samples one would expect linewidths in the range of several tens of wavenumbers \cite{Casiraghi2009}. This clearly excludes the interpretation that the observed stiffening of the G mode in the antidot sample is caused by disorder. The narrow and almost unchanged linewidths indicate that doping is the reason for the upshift, caused by the nonadiabatic removal of the Kohn anomaly (see below) \cite{Pisana2007, Yan2007}. The most important results of the present work are shown in Fig.\ \ref{Fig3}. Figure \ref{Fig3}a displays a summary of the most relevant data from the whole series of investigated antidot samples. If we plot, sample by sample, the positions of the observed 2D modes versus the positions of the G modes, we find an approximately linear correlation (Fig.\ \ref{Fig3}a). The linewidths of the observed G modes in the different samples tentatively decrease with increasing G mode energy (inset of
Fig.\ \ref{Fig3}a). The correlation between G mode and 2D mode positions was recently investigated systematically on a large number of graphene samples, both, in as-deposited and in gated samples \cite{Casiraghi2009, Das2008}. It was reported that while the 2D mode position {\em decreases} with increasing G mode position in n-doped samples, it {\em increases} with increasing G mode position in p-doped graphene \cite{Casiraghi2009}. The dashed line in Fig.\ \ref{Fig3}a indicates, for the range of G mode energies relevant for our work, the slope of the approximately linear correlation, as reported in Ref.\ \cite{Casiraghi2009} for p-doped graphene. In stark contrast, for n-type doping the slope should be negative. One can see a fairly good agreement of the data, extracted from our measurements on the antidot samples, with the dashed line in Fig.\ \ref{Fig3}a. Please note that the 2D mode energies in our experiments are slightly lower than in Ref.\ \cite{Casiraghi2009}, since we measured with a 532 nm line (514 nm line in Ref.\ \cite{Casiraghi2009}): The 2D mode energy increases with decreasing laser wavelength due to the doubly resonant intervalley scattering process \cite{Ferrari2006, Malard2007}. Therefore, the dashed line in Fig.\ \ref{Fig3}a is rigidly shifted by about 17 cm$^{-1}$ to lower energies as compared to 2D mode energies in Fig.\ 3 of Ref.\ \cite{Casiraghi2009}. Obviously, our data in Fig.\ \ref{Fig3}a suggest a p-type doping of the antidot lattices. We note here, that in previous Raman investigations of graphene edges, also indications of a stronger doping in close vicinity to the edges, as compared to the bulk of the samples, were found \cite{Casiraghi2009a}. Based on this, and on our observations, we suspect that the doping effect reaches up to a maximum distance of a few tens of nanometers around the edges of the holes. Hence, the amount of p doping appears to increase systematically with the ratio of etched to unetched graphene area, which is tentatively verified by the data in Fig.\ \ref{Fig3}a. As a result, the plain reference sample regions are not affected by the doping. In order to quantify the amount of p doping, we have compared our results to measurements on an un-patterned gated sample, where doping can be introduced electrostatically (cf.\ Refs.\ \cite{Pisana2007, Yan2007}). Figure \ref{Fig3}b shows the observed G mode positions and linewidths versus applied gate voltage for electrostatic p doping. The gate voltage where the Fermi energy is at the Dirac point is indicated by a vertical blue dashed line. The G mode is Landau-damped due to decay into vertical electron-hole excitations, and hence has a larger linewidth, as long as the Fermi energy is smaller than half of the G mode energy (gray-shaded region in Fig.\ \ref{Fig3}b) \cite{Yan2007}. In accordance with Ref.\ \cite{Yan2007}, the observed G mode energy in the gated sample increases almost linearly with increasing negative gate voltage (indicated by dashed black line in Fig.\ \ref{Fig3}b) due to the nonadiabatic removal of the Kohn anomaly at the $\Gamma$ point \cite{Lazzeri2006}. In the low-density range, within the Landau-damped region, the disorder potential dominates the charge state \cite{Stampfer2007}, leaving a nonzero average doping, and, hence, an almost constant G mode energy. From transport experiments on similar samples we know that the relation between hole (electron) density and applied gate voltage, measured with respect to the Dirac point, is $p\sim7.2\times 10^{10}$ cm$^{-2}\times$ V$_{\rm gate}$ [V]. From this and the measured slope of G mode energy versus gate voltage in Fig.\ \ref{Fig3}b, we conclude that the G mode energy increases by 1 cm$^{-1}$ in the graphene samples per $\sim 4.5\times 10^{11}$ cm$^{-2}$ increase in hole density, starting at a G mode energy of 1580 cm$^{-1}$ at zero density. Of course, this relation holds for densities $>1\times 10^{12}$ cm$^{-2}$, above the disorder-dominated regime, only. With this calibration we can assign hole densities between about $3\times 10^{12}$ cm$^{-2}$ (400 nm sample in Fig.\ \ref{Fig3}a) and $7\times 10^{12}$ cm$^{-2}$ (80 nm sample in Fig.\ \ref{Fig3}a) to our investigated antidot samples in Fig.\ \ref{Fig3}a.

In conclusion, we have investigated antidot graphene samples by scanning Raman spectroscopy. The overall crystalline quality of the antidot lattices is as good as that of the as-deposited layers. An observed systematic stiffening of the G mode in the antidot samples, accompanied by a similar increase in 2D mode energy lead to the conclusion that the introduction of the antidot lattices causes a p-type doping. In all likelihood, the observed p doping is a consequence of the patterning process here, and not a result of the specific antidot pattern.

We acknowledge financial support by the DFG via GRK 1570 and SPP 1459.

\end{document}